# The First Room-Temperature Ambient-Pressure Superconductor

Sukbae Lee[1*], Ji-Hoon Kim[1], Young-Wan Kwon[2†]

[1]Quantum Energy Research Centre, Inc., (Q-centre, Inc.), B1, 46-24, Songi-ro 23 gil, Songpa-gu, Seoul 05822, Korea

[2]KU-KIST Graduate School of Converging Science and Technology, Korea University, Seoul 02841, Korea

## Abstract

For the first time in the world, we succeeded in synthesizing the room-temperature superconductor ($T_c \geq 400$ K, 127 °C) working at ambient pressure with a modified lead-apatite (LK-99) structure. The superconductivity of LK-99 is proved with the Critical temperature ($T_c$), Zero-resistivity, Critical current ($I_c$), Critical magnetic field ($H_c$), and the Meissner effect. The superconductivity of LK-99 originates from minute structural distortion by a slight volume shrinkage (0.48 %), not by external factors such as temperature and pressure. The shrinkage is caused by $Cu^{2+}$ substitution of $Pb^{2+}$(2) ions in the insulating network of Pb(2)-phosphate and it generates the stress. It concurrently transfers to Pb(1) of the cylindrical column resulting in distortion of the cylindrical column interface, which creates superconducting quantum wells (SQWs) in the interface. The heat capacity results indicated that the new model is suitable for explaining the superconductivity of LK-99. The unique structure of LK-99 that allows the minute distorted structure to be maintained in the interfaces is the most important factor that LK-99 maintains and exhibits superconductivity at room temperatures and ambient pressure.

*stsaram@qcentre.co.kr

†ywkwon@korea.ac.kr

## Introduction

Since the discovery of the first superconductor(*1*), many efforts to search for new room-temperature superconductors have been carried out worldwide(*2, 3*) through their experimental clarity or/and theoretical perspectives(*4-8*). The recent success of developing room-temperature superconductors with hydrogen sulfide(*9*) and yttrium super-hydride(*10*) has great attention worldwide, which is expected by strong electron-phonon coupling theory with high-frequency hydrogen phonon modes(*11, 12*). However, it is difficult to apply them to actual application devices in daily life because of the tremendously high pressure, and more efforts are being made to overcome the high-pressure problem(*13*).

For the first time in the world, we report the success in synthesizing a room-temperature and ambient-pressure superconductor with a chemical approach to solve the temperature and pressure problem. We named the first room temperature and ambient pressure superconductor LK-99. The superconductivity of LK-99 proved with the Critical temperature ($T_c$), Zero-resistivity, Critical current ($I_c$), Critical magnetic field ($H_c$), and Meissner effect(*14*, *15*). Several data were collected and analyzed in detail to figure out the puzzle of superconductivity of LK-99: X-ray diffraction (XRD), X-ray photoelectron spectroscopy (XPS), Electron Paramagnetic Resonance Spectroscopy (EPR), Heat Capacity, and Superconducting quantum interference device (SQUID) data. Henceforth in this paper, we will report and discuss our new findings including superconducting quantum wells associated with the superconductivity of LK-99.

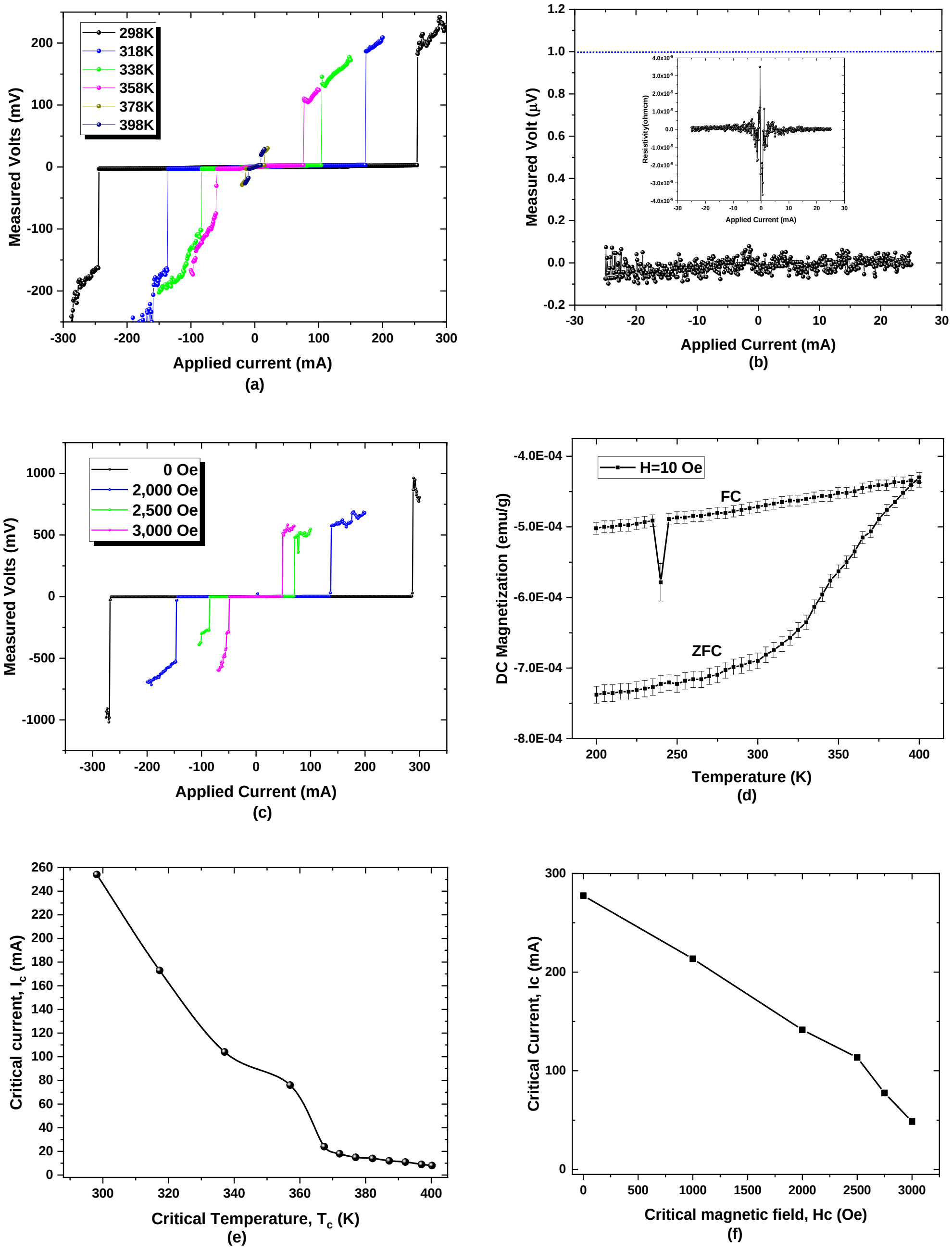

Figure1. (a) Measured Voltage vs. Applied Current at various Temperature, (b) Zero Resistivity with thin film, (c) Measured Voltage vs. Applied Current under various External Magnetic field (Oe), (d) DC magnetization of Field-Cooling and Zero Field-Cooling with 10 Oe magnetic field, (e) Critical current vs. Critical temperature, (f) Critical current vs. Critical magnetic field.

Figure 1(a) shows the measured voltage vs. applied current at various temperatures (298 K ~ 398 K), the measurement for Figure 1(a) was performed with direct current (DC) polarity change by each 20 K increment of temperature in the vacuum of $10^{-3}$ *Torr*. In various bulk samples, specific resistance was measured in the range of $10^{-6}$ to $10^{-9}$ $\Omega\cdot$cm. Also, a thin film of LK-99 was fabricated on the precision glass plate by a thermal vapor deposition process (UNIVAC, Korea). Figure 1(b) shows the Zero-resistivity of the thin film of LK-99, which satisfied the zero-resistivity of international standards(*16, 17*) as a new superconductor. According to the International Electrotechnical Commission standards(*16*), there are two equivalent criteria(*17*) for superconductivity: the electric field criterion with 1 μV/cm or 0.1 μV/cm and resistivity criterion with $10^{-11}$ $\Omega\cdot$cm. As shown in Figure 1(b), the measured voltage was obtained in the range of 0.1 μV/cm during the applied current increasing and decreasing. Also, the resistivity was calculated in the order of $10^{-10}$ ~ $10^{-11}$ $\Omega\cdot$cm. As the grain boundary is decreased, the residual resistance of the thin film decreased(*18*). Figure1(c) shows the external magnetic field (H) dependence of applied current. Even up to 400 K, the DC magnetization value of Zero field cooling and Field cooling with 10 Oe was still negative in Figure 1(d). These results indicate that the superconducting phase still exists under 10 Oe up to 400 K. Additionally, the critical current value was not yet zero (7 mA) even at 400 K and 3000 Oe or more in Figure 1(e) and (f). Therefore, we judge that the critical temperature of LK-99 is over 400 K. Moreover, since LK-99 has a polycrystalline morphology shown in Figure 2, the non-uniform resistivity of the bulk samples can be explained as inter-grain boundaries, intergranular vortex flow, free vortex flow(*18-20*) of polycrystalline superconducting phase. The Josephson-like phenomenon (Figure S1(a) in supplementary materials) for the under-damped junction of superconductor-normal metal-superconductor(*21, 22*) or Inter-grain coupled superconductors(*23*) and the thermoelectric effect(*24-26*) of the inter- or intra-grain network(Figure S1(b) in supplementary materials) were also observed.

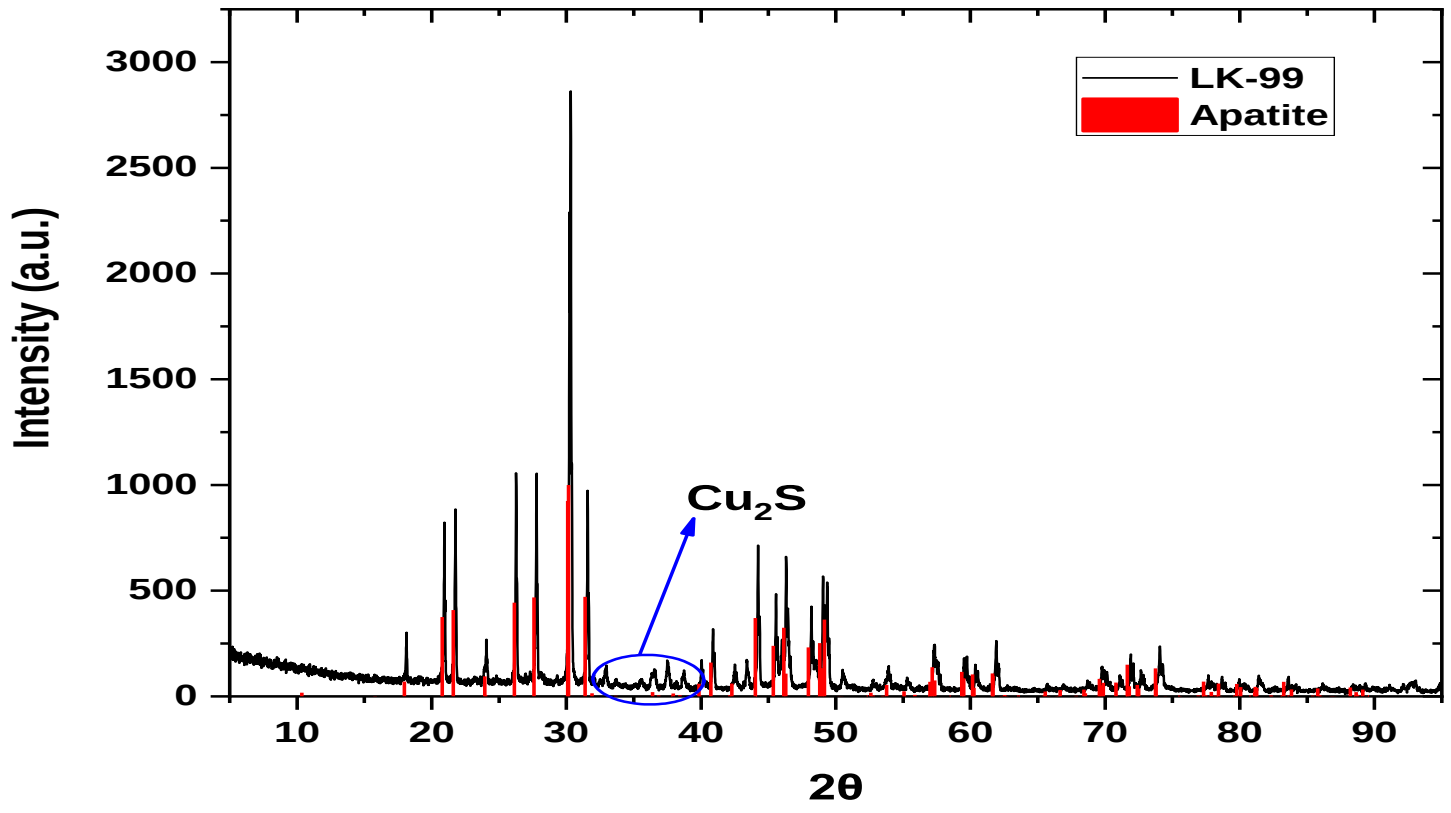

Figure 2. XRD results of LK-99 matched with COD, the original XRD data was only Kα2-stripped without any other processing.

Figure 2 shows the results of LK-99 matched with the QualX2.0 software(*27*) and proved by simulated data using VESTA software(*28*), which uses Crystallography Open Database (COD) to perform the search-match operation(*27*). It shows that LK-99 is polycrystalline. The main peaks were well matched with the lead-apatite (AP) structure, and few $Cu_2S$ impurities also shown. The crystal system of the original lead-apatite is hexagonal ($P6_3/m$, 176) with the cell parameters a=9.865 Å and c=7.431 Å. However, LK-99 shows a slight shrinkage compared to the Lead-Apatite with parameters of a=9.843 Å and c=7.428 Å. The volume reduction of LK-99 is 0.48 %.

Humankind has long learned that the properties of matter stem from its structure. However, so far, the correlation between superconductivity and the structural change of material has hardly been properly clarified. In fact, the two main factors affecting the generation of superconductivity of superconductors discovered so far are temperature and pressure(*29, 30*). And both temperature and pressure affect its volume. It seems that the stress generated by the decrease in volume under the low temperature or high pressure causes a minute strain or distortion. Although it is difficult to observe the minute structural changes in superconducting materials, this structural change seems to bring the superconductivity of it.

Representative examples showing the superconductivity due to the external factors (pressure and temperature) and the internal factors (oxygen doping) are CuO-based superconducting materials. The structure of CuO changes from the tetragonal structure to the orthogonal structure depending on the oxygen doping contents. In addition, the higher $T_c$ is observed under the high pressure condition(*29*). In the CuO based superconductors, it may be thought that the stress generated by the inconsistency of the amount of change in the c-axis and a and b-axis affects the intermediate CuO layer, causing structural distortion, and superconductivity because the changes in the c-axis seems to be the most sensitive to changes in temperature and pressure. And other examples are what we think of as the appearance of superconductivity caused by distortion or strain caused by fine physical stress. When the FeSe monolayer on the top of the insulator is stacked, the highest $T_c$ achieves up to 65 K(*31*) and even up to 109 K(*32*). However, the $T_c$ lowers to 10 K as layer stacks increases (*33*-*35*). In the case of CuO monolayer films on $Bi_2Sr_2Cu_2O_{8+\delta}$, Zhong et al. identified two distinct and spatially separated gaps: U-like gap and V-like gap(*36*). In 2019, Choi et al. proved the effect of 3D strain for $T_c$ in the cuprate system(*37*), and $BaFe_2As_2$ system also showed strain-induced superconductivity(*38*) in 2013. In 2008, Gozar et al. reported high-temperature interface superconductivity between metallic and insulating copper oxides(*39*). The thinner the layer, the greater the stress-inducing effect, the greater the strain, which seems to be the higher the superconducting transition temperature. Therefore, we argue that the stress caused by temperature and pressure brings a minute structural distortion and strain, which create an electronic state for superconductivity.

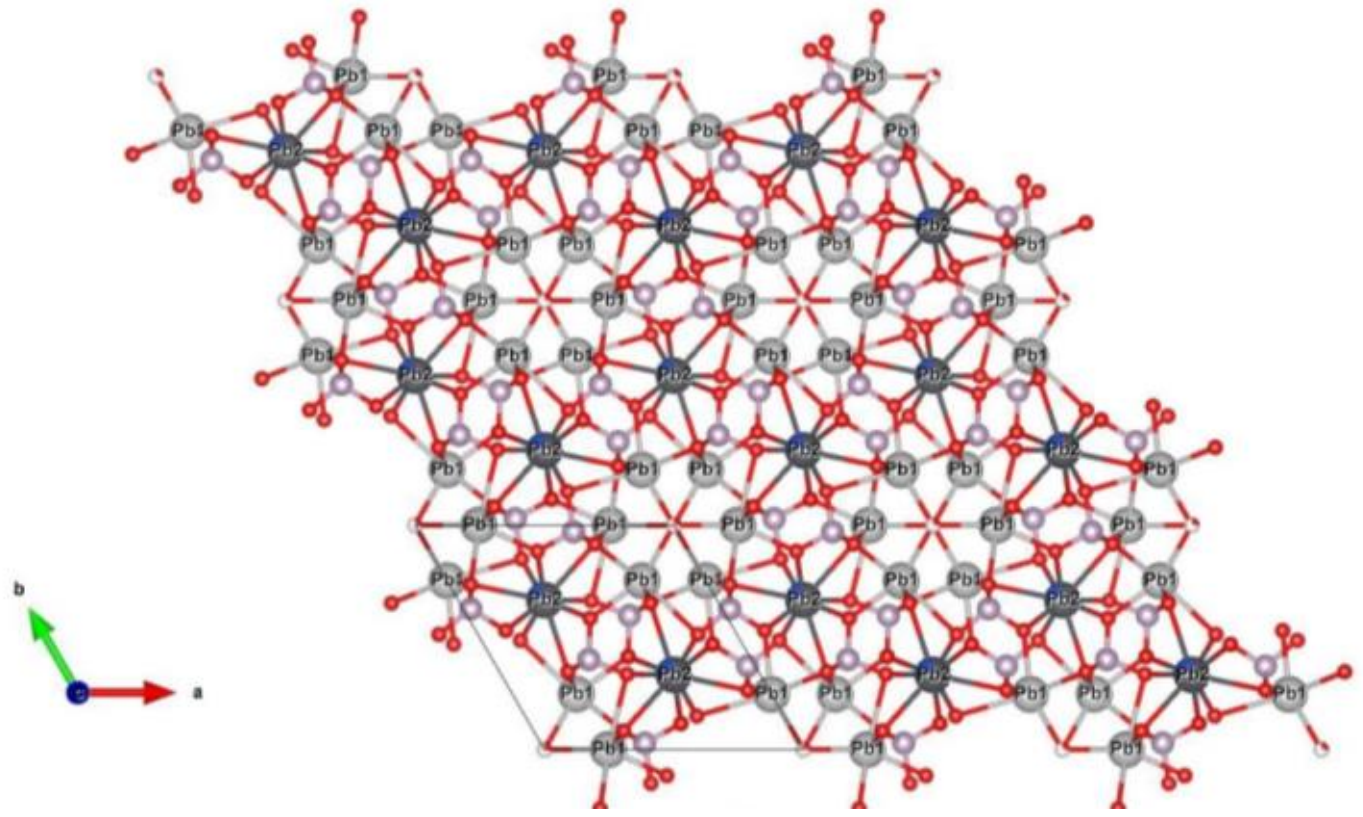

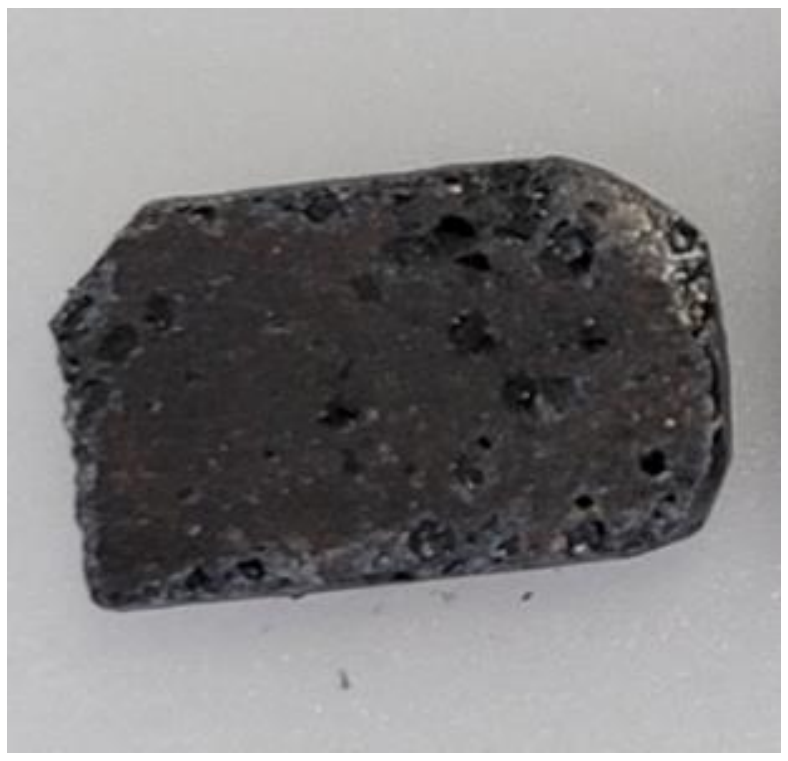

**(a)** **(b)**

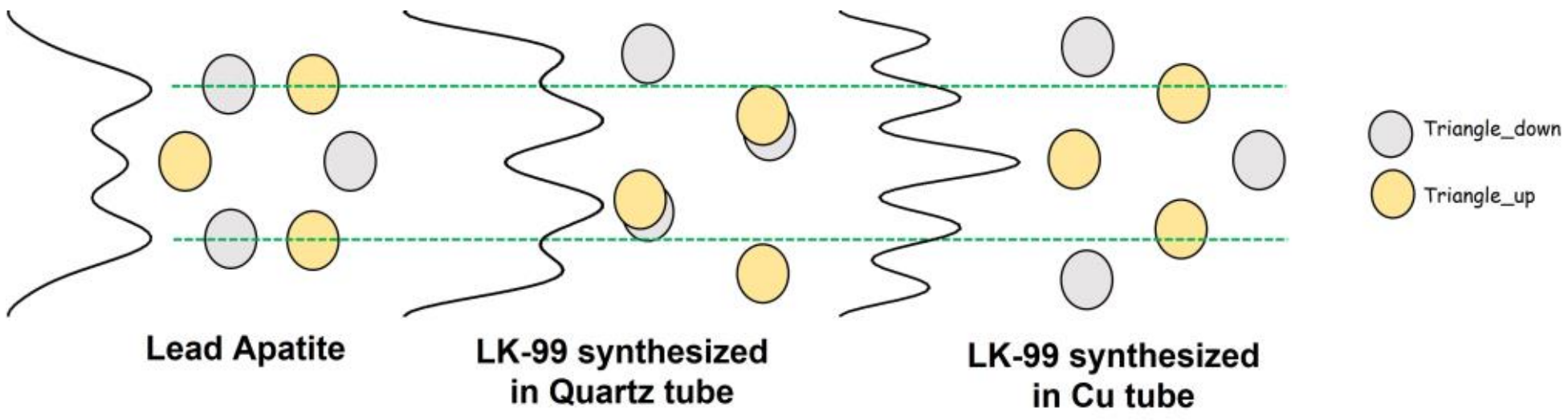
Lead Apatite
LK-99 synthesized in Quartz tube
LK-99 synthesized in Cu tube
Triangle_down
Triangle_up

**(c)**

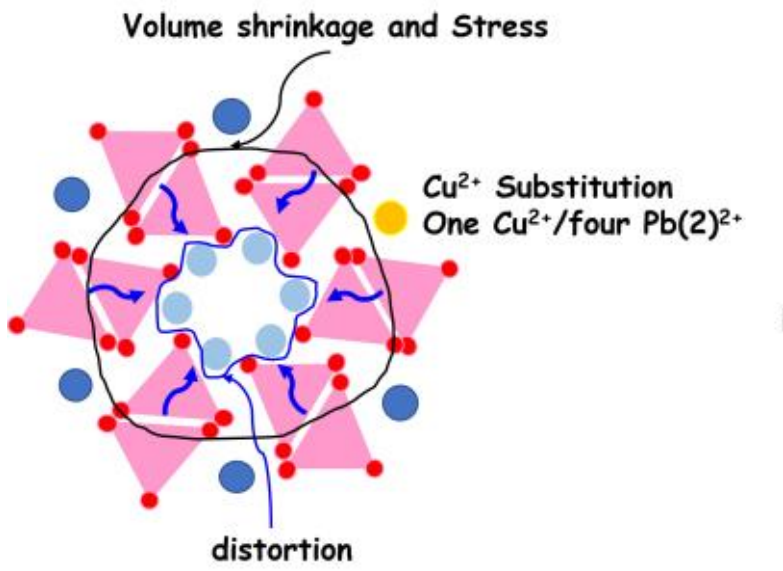
Volume shrinkage and Stress
Cu2+ Substitution
One Cu2+/four Pb(2)2+
distortion

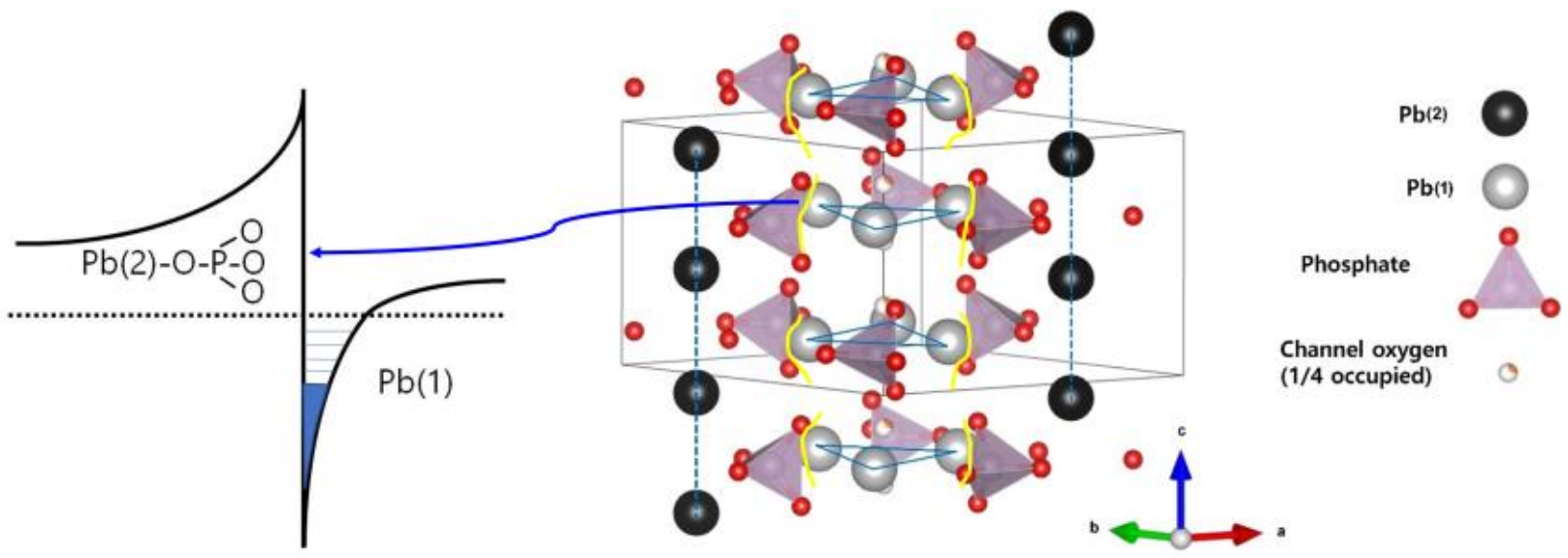

**(d)** **(e)**

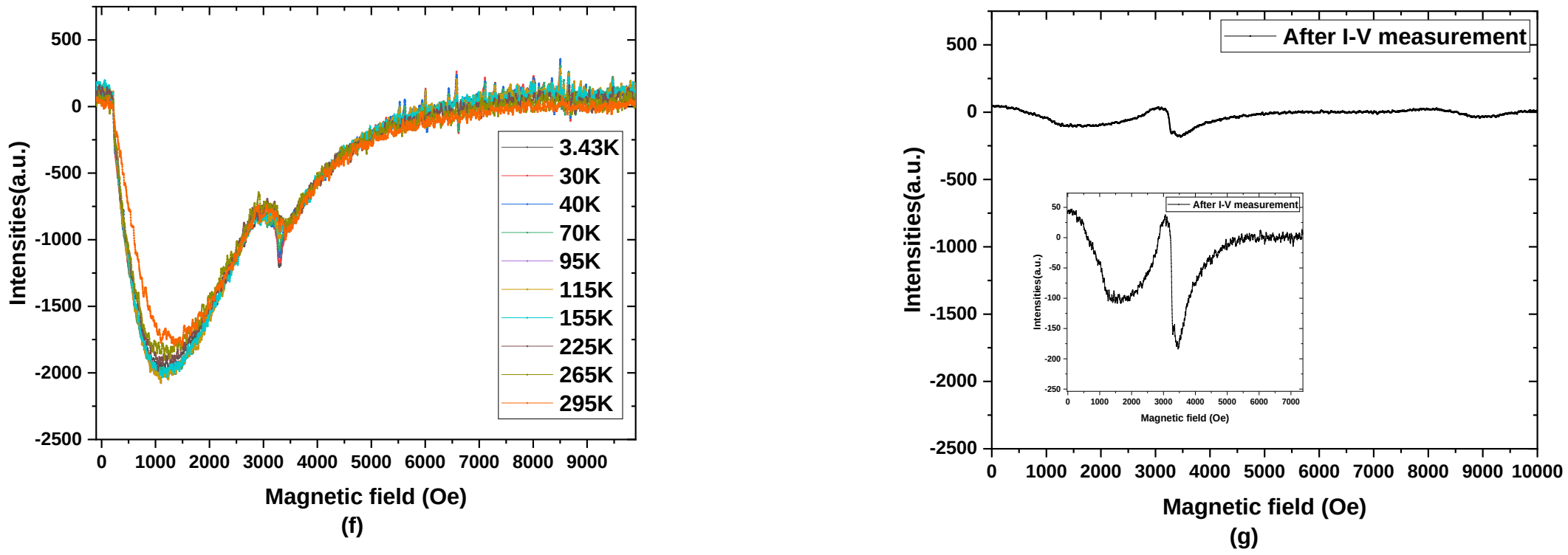

Figure 3. (a) the top-view of c-axis of Lead Apatite and LK-99 produced by VESTA software(*28*), (b) the synthesized LK-99, (c) the calculated electron density map of Pb(1) of cylindrical column interface of Lead Apatite and LK-99 synthesized with the different tubes (Cu and Quartz tube), (d) the top-view of the stressed interface of cylindrical column produced by VESTA software, (e) the side-view of the cylindrical column(*28*) and the predicted superconducting quantum well, (f) EPR signals of LK-99 from 3.43 K to 295 K, (g) EPR signal of LK-99 after Current-Voltage measurement.

LK-99 is a gray-black color, as shown in Figure 3(b). It is the superconductor with the same color as typical superconductors. The system of lead-apatite(*40*), $Pb_{10}(PO_4)_6O$, is an ivory-colored powder and an insulator. The LK-99 has a three-dimensional network structure shown in figure 3(a) and a cylindrical column surrounded by an insulating tetrahedral $PO_4$ network structure. And the cylindrical columns arranged at intervals of 6.541 Å are composed of asymmetric polyhedral six Pb(1)-$O_n$ components consisting of two oppositely shaped triangles as shown in the side-view of the cylindrical column in Figure 3(e). The polyhedral four Pb(2) are components of insulating $PO_4$ network structures of LK-99. The modified lead-apatite of LK-99 is expected to be $Pb_{10-x}Cu_x(PO_4)_6O$, x=0.9~1.0, in which one of the four Pb(2) ions approximately has been replaced by Cu(II) ion at the polyhedral Pb(2) sites of the lead-apatite(*41*). In LK-99, the ratio of copper was determined based on the atomic % data of XPS and confirmed by the Debye heat capacity model with heat capacity result and chemical formula $Pb_{10-x}Cu_x(PO_4)_6O$, x=0.9~1.0. And it is well consistent with previous published research result(*41*). This replacement of $Cu^{2+}$ ions in LK-99 resulted in a volume reduction of 0.48 % because $Cu^{2+}$ ions (87 pm) are smaller than $Pb^{2+}$ ions (133 pm). The stress was occurred in the network portion and then affected the appearance of superconductivity.

As a result of more precise analysis of XRD and XPS data, it reveals where the stress from volume reduction was finally affected. For the purpose of determining the variation of Pb(1) positions, One-dimensional electron density calculation(*42*) along one crystallographic axis via the Fourier transform of the calculated structure factor was used. The electron density was calculated along the z-direction, ρ(c), based on the (00*l*) reflection intensities of XRD data using the following equation.

$$\rho(z/c) = \frac{1}{c}\sum_{l=-\infty}^{\infty} F(00l)e^{-i\left(\frac{2\pi l z}{c}\right)}$$

Where *l*, F(00*l*), c and z are the order of the (00*l*) diffraction peak, the structure factor, the unit cell parameter of the c-axis, and the atomic coordinate along the z-axis, respectively. Since LK-99 has a hexagonal structure, the above equation was applied for the electron density calculation of Pb(1) along the z-direction (c-axis) and x-direction (a-axis), and ρ(c) and ρ(a), based on the (00*l*) and (*h*00) reflection intensities of XRD data in Figure 2. As the result, the position of Pb(1) constituting the cylindrical column is slightly shifted in the a-axis plane from the original position by the substitution of $Cu^{2+}$ inward or outward in Figure 3(c). In the repeated triangular structure of Pb(1) of the cylindrical column, the distance between Pb(1) in one layer is decreased to 2.61815 Å, and the next layer is increased to 5.23476 Å from the original distance of 3.03340 Å. However, the distance (3.7140Å) through the c-axis between the triangular layers of Pb(1) of LK-99 remains almost unchanged from lead-apatite (3.7153Å). According to the analyzed results of XPS data, the binding energies (B.E.) of Pb(2) and Phosphorus were unchanged. Although, the tetrahedral Phosphorus splitting value between $2p_{3/2}$ and $2p_{1/2}$very slightly increased from 0.68 eV to 0.69 eV, and all oxygens B.E. are quite a bit increased by 0.21 eV, 0.33 eV, 0.56 eV, 0.89 eV, respectively. Also, the B.E. value of Pb(1) is slightly decreased by 0.03 eV. From the results described above, it can be seen that the volume reduction was caused by the replacement of $Cu^{2+}$ ions, and the stress caused by the volume reduction caused the position change of Pb(1) and the binding energy change between oxygen atoms adjacent to Pb(1). Figure 3(f) shows the EPR signal of LK-99. It is the same as the heterojunction quantum well such as Si/SiGe(*43*), natural DNA(*44*) in a dry state, and $\alpha$-$Fe_2O_3$(*45*) doped with $Mg^{2+}$. The EPR signals were interpreted as a cyclotron resonance signal of 2-dimensional electron gas (2-DEG) of quantum well, which also confirmed the creation of quantum wells in the interface between Pb(1) and Phosphate of LK-99. In addition, the substituted $Cu^{2+}$ ions were appeared at 3000 Oe in EPR signals as shown in Figure 3(f). The superconductivity with the 2-DEG system of GaAs/AlGaAs(*46*) and DNA(*47*) was reported at 0.3 K and 1 K. The superconductor having interface structure like a heterojunction of the 2-DEG system such as $LaAlO_3$/$SrTiO_3$(*48-50*) has been shown the superconductivity. Hence, it is confirmed that the superconducting quantum well (SQW) was generated between Pb(1) and Oxygens of Phosphate by structural distortion and strain. The predicted SQW was illustrated in Figure 3(e). The superconductivity of LK-99 is deeply related to SQW. In 2002, Koji et al. already reported the similar EPR signals as shown in Figure 3(f) and the superconductivity with $RuSr_2GdCu_2O_8$ system(*51*). However, they did not interpret the EPR signals as SQW and to associate it with the appearance of superconductivity.

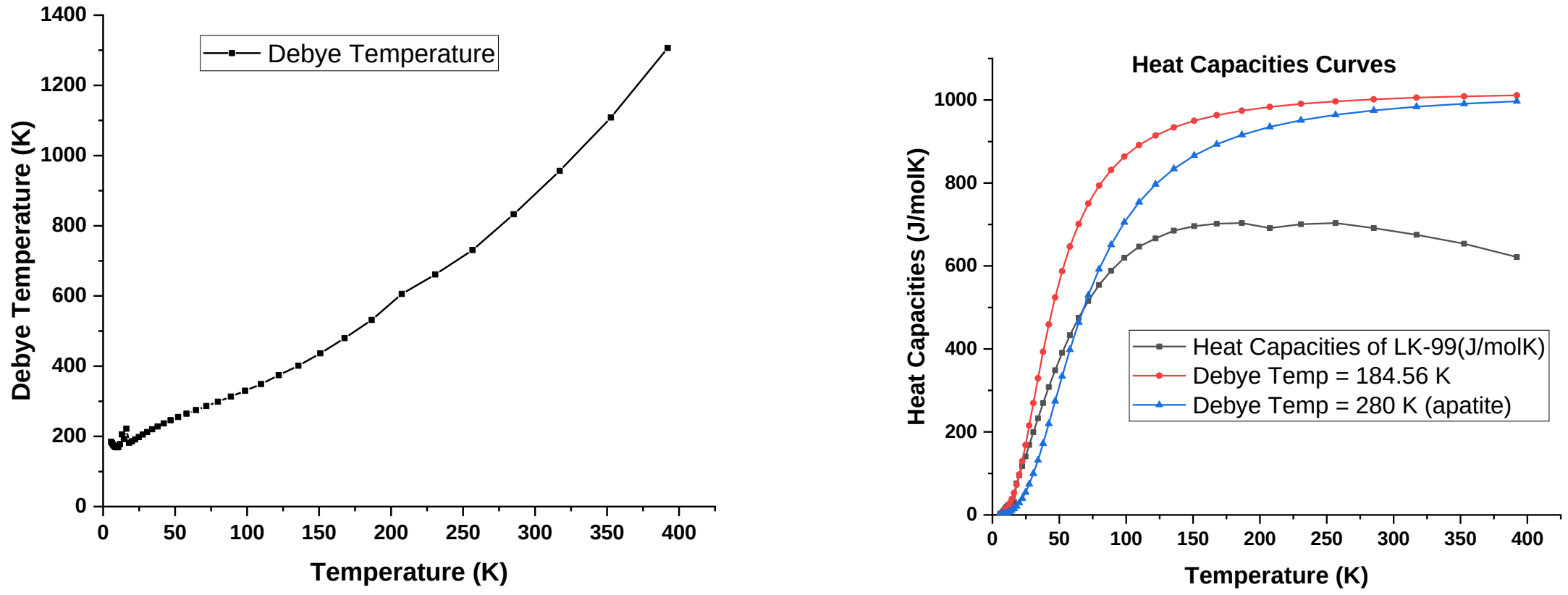

Figure 4. (a) Calculated Debye Temperature from Heat Capacity data, (b) Heat Capacity curves of LK-99.

The heat capacity of LK-99 was measured from 5 K to 400 K using Physical Property Measurement System (PPMS, DinaCool-9, Quantum Design, USA), and Debye temperature was calculated based on the heat capacity data using $Pb_{10-x}Cu_x(PO_4)_6O$ formula (x = 1) through the following Debye heat capacity equation;

$$C_v = 9rNk\frac{T^3}{\theta^3}\int_0^{\theta/T}\frac{x^4e^x dx}{(e^x-1)^2}$$

where $C_v$ is heat capacity, $r$ is the number of atoms per molecule, $N$ is the number of molecules, and $k$ is the Boltzmann constant. Figure 4 shows that the Debye temperature of LK-99 is constantly changing from about 184 K to 1300 K. Thus, $T_c$ cannot be calculated with the electron-phonon prediction models. The blue line curve (Debye temperature = 280 K) is the calculated heat capacity of the typical apatite because the Debye temperature of the typical apatite is 280 K. The red line curve is the calculated heat capacity results based on Debye temperature (184.56 K) at a low temperature (5 K) of LK-99. In the case of conventional superconductors, it is relatively well explained by the electron-phonon model. And in the electron-phonon model, one of the most influential parameters of predicting $T_c$ is Debye temperature(*52*). The heat capacity curve (black line curve) of LK-99 does not follow the Debye models, since the normal vibrational mode is limited by $Cu^{2+}$ ions replacement in the network part of the LK-99. This result also confirms that LK-99 has a structure distorted by the substitution of $Cu^{2+}$ ions. In terms of the EPR experimental results shown in LK-99, the expression of superconductivity can be explained by the formation of SQWs. There is also a paper explaining superconductivity with the quantum well model(*36*). However, they did not show EPR signals like ours. Figure 3(d) is the EPR signal measured on the sample (LK-99) after Current-Voltage measurement. Compared to the signal in which no current is applied (Figure 3(f)), the overall signal intensities decreased, and the signal intensity of cyclotron resonance was relatively decreased in Figure 3(g). Additionally, in case of CuO based superconductors, the large absorption EPR signal appeared at a very low temperature and below 1000 Oe external magnetic field(*53*, *54*), and we also observed with YBCO and Bi2212. The signal below 1000 Oe external magnetic field can be interpreted as a signal by superconducting electrons(*53*, *54*).

Josephson installed a thin insulator between superconductors and discovered tunneling through which current flows. Tunneling occurred between the superconductors(*55*). Likewise, if electrons are moved by tunneling between SQWs, the resistance will naturally be zero. Tunneling between SQWs is likely to be possible, as SQWs are expected to be formed at 3.7 Å ~ 6.5 Å intervals in LK-99. According to Kim's paper(*52*), he used 1250 K Debye temperature for his calculation of room-temperature superconductor, which is mentioned for hydride in 1968 by Ashcroft(*56*). Even in that case, he argued that the electron-electron interaction contributes to the higher temperature superconductivity than the electron-phonon interaction. In LK-99 system, the applied current seems to be transported through the correlated SQWs of the cylindrical column via the tunneling process and coherently transported in 3-dimension with all SQWs together. The additional experimental results and discussions on LK-99 will be published immediately in the next paper, including an interesting controllable levitation phenomenon and the coexistence of magnetism and superconductivity, theoretical calculation, etc. LK-99 will be an attractive substance for many researchers that can solve various confusing puzzles such as the pairing process and coexistence with magnetism, etc., related to superconductivity.

Consequently, why does LK-99 exhibit superconductivity at room temperature and ambient pressure? This is because the stress generated by the $Cu^{2+}$ replacement of $Pb(2)^{2+}$ ion was not relieved due to the structural uniqueness of LK-99 and at the same time was appropriately transferred to the interface of the cylindrical column. In other words, the Pb(1) atoms in the cylindrical column interface of LK-99 occupy a structurally limited space. These atoms are entirely affected by the stress and strain generated by $Cu^{2+}$ ions. Therefore, SQWs can be generated in the interface by an appropriate amount of distortion(*57*) at room temperature and ambient pressure without a relaxation. From this point of view, the stress due to volume contraction by temperature and pressure is relieved and disappeared in CuO- and Fe-based superconductor systems because the relaxation process cannot be limited due to the structural freedom. Therefore, they need an appropriate temperature or pressure to limit the structural freedom and to achieve the SQW generation. The LK-99 is a very useful material for the study of superconductivity puzzles at room temperature. All evidence and explanation lead that LK-99 is the first room-temperature and ambient-pressure superconductor. The LK-99 has many possibilities for various applications such as magnet, motor, cable, levitation train, power cable, qubit for a quantum computer, *THz* Antennas, etc. We believe that our new development will be a brand-new historical event that opens a new era for humankind.

**Acknowledgments:** We thank Prof. Emeritus Keun Ho Auh (In 1986, his surname was indicated as Orr in the paper(*41*) at the time), Dr. Ali. C. Basaran and Dr. Juan Pereiro Viterbo for fruitful communications and advice. We also thank to Se Woong Ki and Byung Kyu Lee for the financial support of Quantum Energy Research Centre, Inc., (Q-centre, Inc.). Finally, we thanks to Sungyeon Im and SooMin An for various support.

**Funding:** This research was mainly supported by Quantum Energy Research Centre, Inc. and was also partially supported by Basic Science Research Program through the National Research Foundation of Korea (NRF) funded by the Ministry of Education(2019R111A1A01059675) and Young-Wan Kwon is supported by a Korea University Grant.

**Author contributions:** Sukbae Lee conceived, organized, and led the project. Ji-Hoon Kim was mainly responsible for LK-99 synthesis. He developed the synthetic method of LK-99 through the reaction mechanism study and XRD data analysis. Sukbae Lee, Ji-Hoon Kim and Young-Wan Kwon contributed XRD, XPS, EPR, MPMS, PPMS, and Electrical data collection. Young-Wan Kwon proposed an ESR method to find superconducting materials and provided ESR equipment. He interpreted the superconductivity of LK-99 through the analysis of all the data and wrote the manuscript. Sukbae Lee, Ji-Hoon Kim and Young-Wan Kwon discussed all of results and reviewed manuscript.

**Competing interests:** “Authors declare that they have no competing interests.”

**Data and materials availability:** "All data are available in the main text or the supplementary materials."

# Supplementary Materials

## Materials and Methods

### Sample Synthesis and Preparation

For the sample synthesis of the LK-99, the general solid-state reaction was used. Lanarkite and $Cu_3P$ were uniformly mixed in a molar ratio of 1:1 in an agate mortar with a pestle. The sample was put into a reaction tube, sealed under the vacuum of $10^{-5}$ *Torr* and reacted at 925°C for 10 hr. After the reaction, a dark gray ingot was obtained reproducibly and then made into the shape of thin cuboids for electrical measurements. For other analyses, it was pulverized and used as the form of powder. The reagents used for the above solid- state reaction was PbO (JUNSEI, GR), $PbSO_4$ (KANTO, GR), Cu (DAEJUNG, EP), and P (JUNSEI, EP).

1) Preparation for lanarkite, $Pb_2SO_5$

PbO and $PbSO_4$ powders were uniformly mixed in a molar ratio of 1:1 in an agate mortar with a pestle. And then, after the sample was transferred to an alumina crucible, it was reacted at 725 °C for 24 hour in a furnace. After completion of the reaction, white sample was obtained. It was pulverized with the mortar.

2) Preparation for copper phosphide, $Cu_3P$

Cu and P powders were mixed in each composition ratio. And the sample was transferred in a quartz tube. The tube was sealed under the vacuum of $10^{-5}$ *Torr* and reacted at 550 °C for 48 hour in a furnace. After taken out from the tube, a dark gray ingot was obtained and pulverized.

## Set-up for electrical properties measurement

4-point probe measurement was performed with 4 four pogo test probes (YoungJinSa, Inc., Korea) insulated from each other and arrayed in one line with the same distance (1.2mm). The Keithley 228A and Keithley 182 were used as voltage/current source and sensitive digital voltmeter, respectively. For accurate temperature control and measurement, we developed and used the self-designed heating device with a thermally insulated aluminum mounting plate and halogen lamp, as a heating source. The Self-developed software, using LabView software, was also used for the measurement. All measuring instruments were linked with GPIB interface devices. Temperature measurement was carried out with Keithley 2000 with FLUKE 80BK-DMM K-type thermocouple probe on the sample's surface.

**Collection of XRD data**

XRD data collected with Rigaku (SmartLab, Japan) at Institute of Green Manufacturing Technology Green Manufacturing Research Center of Korea University.

**Measurement of Magnetic Properties with SQUID (MPMS)**

Magnetization measurement is performed with dc mode, 30 mm scan length, 10 scans per measurement, scan time 10s by using Superconducting Quantum Interference Device (SQUID) (MPMS3-Evercool of KAIST Analysis Center for Research Advancement, Korea). Zero field-cooling is processed from 400 K to 200 K without external magnetic field and then the temperature is increased from 200 K to 400 K with 10 Oe magnetic field. Finally, the sample is cooling down from 400 K to 200 K with 10 Oe magnetic field. The 45.814 mg sample is used for this measurement.

**Measurement of Electron Paramagnetic Resonance Spectroscopy**

EPR Spectroscopy is performed over the temperature range from 3.45 K to 295 K using a JES-FA200 ESR X-band spectrometer (Jeol, Japan). The incident microwave power is 0.9980 mW, the receiver gains 100 and the sweep time (1 min). The modulated magnetic field is 10 G at 100 KHz and the swept external magnetic field is -100 ~ 9,900 Oe or 0 Oe ~ 10,000 Oe. The LK-99 sample is set in a 5 mm quartz tube (Wilmad LabGlass, USA), sealed vacuum (5 x $10^{-5}$ torr) for LK-99 before I-V measurement, not sealed vacuum for LK-99 sample after I-V measurement. And then the sealed quartz tube was loaded into a cylinder cavity equipped with a liquid helium cryostat system (Advanced Research Systems, USA).

**Measurement of Heat Capacity by PPMS**

The heat capacity data were collected from 5 K to 400 K with 65.26 mg of LK-99 using by Physical Property Measurement System (PPMS) (DynaCool-9, Quantum Design, USA) in Institute of Next-generation Semiconductor convergence Technology (DGIST, Korea). And the raw data was calibrated with heat capacity references of $Cu_2S$ bulk and nanosheet, which is included in a small amount as an impurity.

**X-ray Photoelectron Spectroscopy (XPS) Study on the lead-apatite and LK-99**

XPS data collected from ULVAC-PHI (X-TOOL, Japan), Institute of Green Manufacturing Technology Green Manufacturing Research Center of Korea University. The XPS data of the Pb, P, O and Cu are deconvoluted using XPSPEAK 4.1 software. The background is corrected using the Shirley background function and the Lorentzian (80%)-Gaussian (20%) sum function is used for fitting.

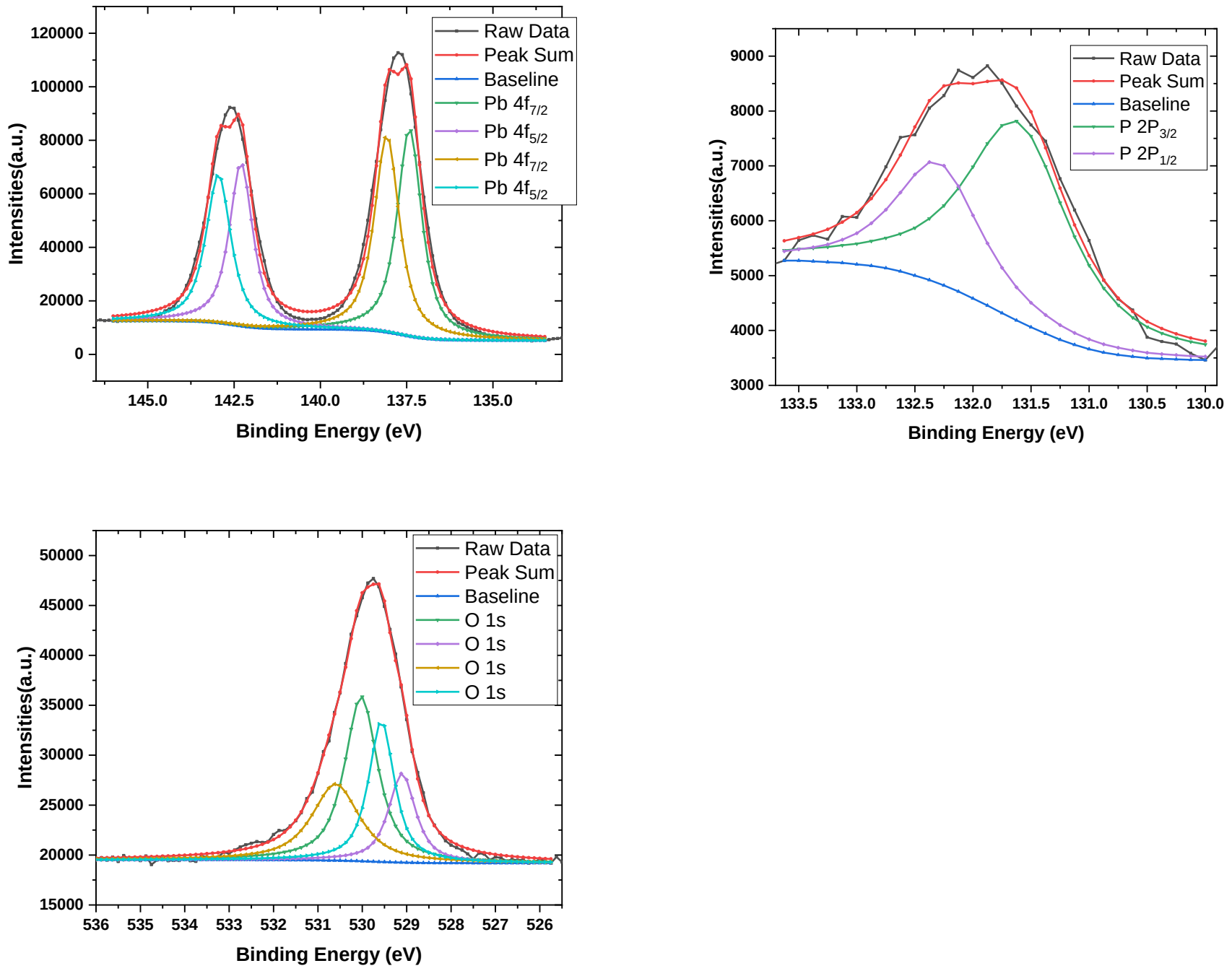

Fig S1(a). the deconvoluted XPS graph Pb $4f_{7/2}$ and $4f_{5/2}$, P $2P_{3/2}$ and $2P_{1/2}$, and O 1s of Lead Apatite ($Pb_{10}(PO_4)_6O$).

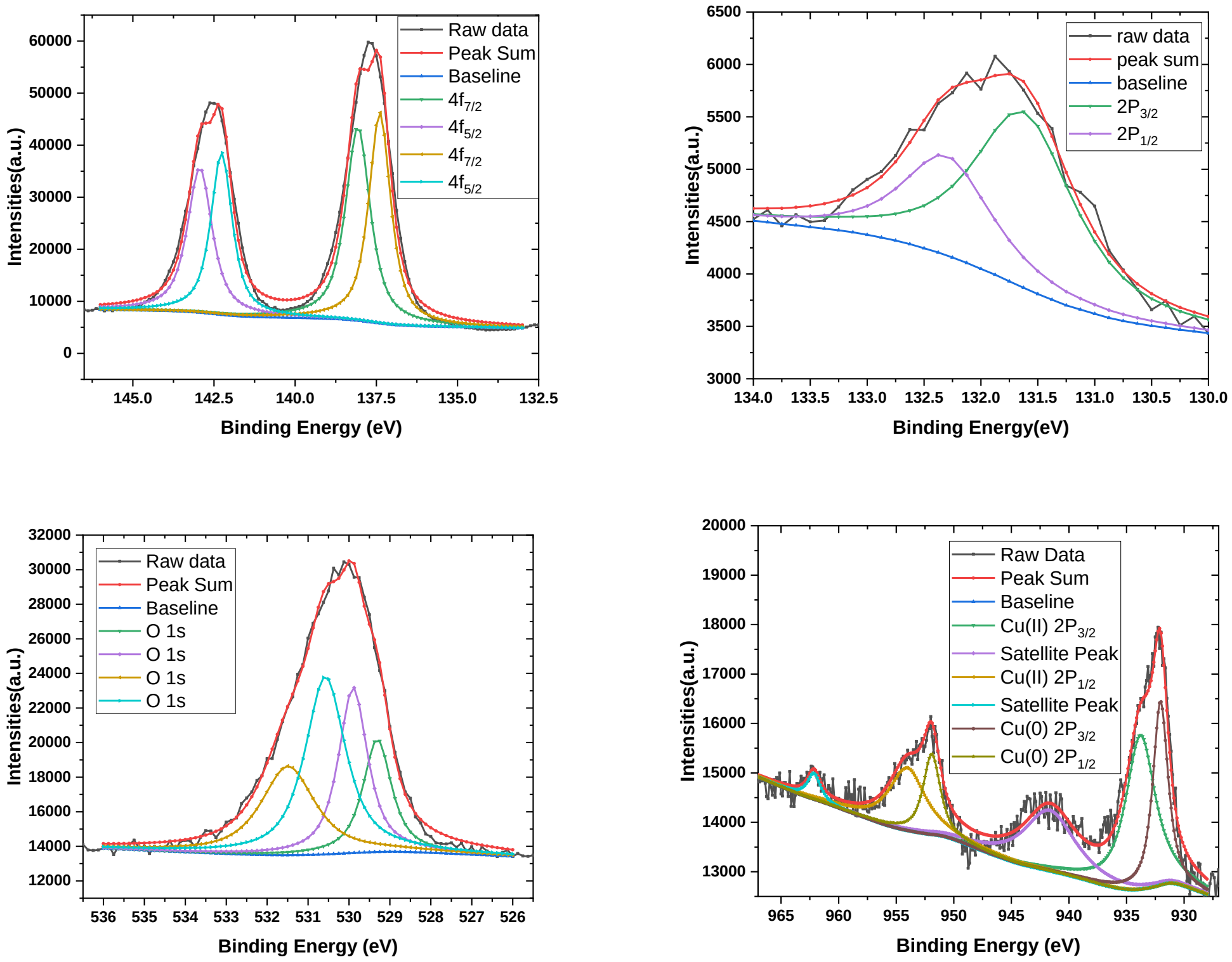

Fig S1(b). the deconvoluted XPS graph Pb $4f_{7/2}$ and $4f_{5/2}$, P $2P_{3/2}$ and $2P_{1/2}$, O 1s, and Cu(II) $2p_{3/2}$ $2p_{1/2}$ of LK-99.

In the XPS results of Cu, Binding Energy (B.E.) 932.05 eV, 951.91 eV is come from Cu(0) atom, which is residue of reactant $Cu_3P$. the only 933.78 eV, 954.03 eV are from the substituted Cu(II) ion of LK-99 matrix, the rest are satellite peaks of Cu(II) ions. And the amount of atomic substitution is calculated using the relative atomic sensitivity. The result is about Cu/Pb=0.9/10.

Table S1. XPS data

| | Pb $4f_{7/2}$, $4f_{5/2}$ | P $2P_{3/2}$ $2P_{1/2}$ | O 1s | Cu $2P_{3/2}$ $2P_{1/2}$ |
|---|---|---|---|---|
| Lead Apatite | Pb(1) 137.42 eV<br>142.3 eV<br>Pb(2) 138.07 eV<br>142.95 eV | 131.61 eV<br>132.3 eV<br>Splitting value: 0.69 eV | O(4) 529.10 eV<br>O(1) 529.57 eV<br>O(2) 530.02 eV<br>O(3) 530.60 eV | |
| LK-99 | Pb(1) 137.39 eV<br>142.27 eV<br>Pb(2) 138.07 eV<br>142.94 eV | 131.62 eV<br>132.3 eV<br>Splitting value: 0.68 eV | O(4) 529.31 eV<br>O(1) 529.9 eV<br>O(2) 530.58 eV<br>O(3) 531.49 eV | Cu(0) 932.05 eV<br>951.91 eV<br>Cu(II) 933.78 eV<br>954.03 eV |

Looking at the table S1, there is a little change in B.E. of Pb(1) among Pb(1) and Pb(2). The four Pb(2) ions is composed of the network structure of apatite, and six Pb(1) ions is composed of the cylindrical column in main text. And four different types of oxygen are bonded to Pb(1) and Pb(2) and made a polyhedral structure like a Figure S3.

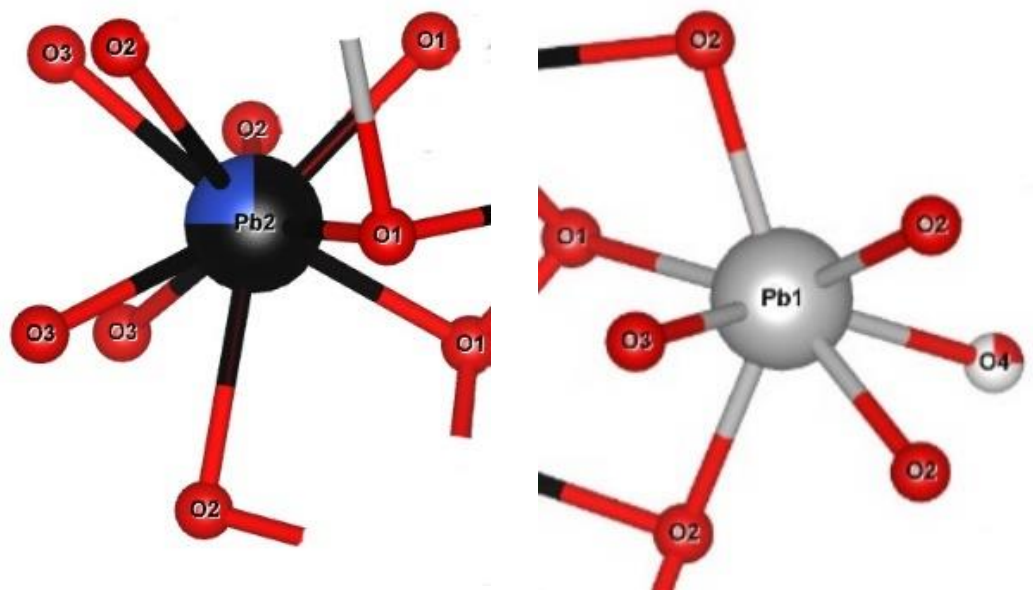

Fig S2. Chemical structures of polyhedral Pb(1)-$O_n$ and Pb(2)-$O_n$